\documentclass[preprint,showpacs,preprintnumbers]{revtex4}
\usepackage{amssymb}
\usepackage{amsmath}
\usepackage{graphicx}
\usepackage{dcolumn}
\usepackage{bm}
\usepackage{color}
\usepackage{subfigure}
\usepackage{epstopdf}
\usepackage{float}

\begin{document}

\title{ Two-photon  Rabi-Stark model}
\author{ Jiong Li$^{1}$ and Qing-Hu Chen$^{1,2,*}$}

\address{
$^{1}$ Department of Physics and Zhejiang Province Key Laboratory of Quantum Technology and Device, Zhejiang University, Hangzhou 310027, China \\
$^{2}$  Collaborative Innovation Center of Advanced Microstructures, Nanjing University, Nanjing 210093, China
 }\date{\today }

\begin{abstract}
The analytically exact solutions to the two-photon Rabi-Stark model are
given by using the Bogoliubov operators approach. Transcendental functions
responsible for the exact solutions are derived. The zeros of the  transcendental functions reproduce completely the regular
spectra. The first-order quantum phase transitions are also detected by the
pole structure of the derived transcendental functions, similar to the
one-photon Rabi-Stark model, different from the two-photon Rabi model where
the energy levels for the ground-state and the first excited
state  do  not cross. The spectra collapse characteristics  resemble  those in two-photon Rabi model.  Some low lying levels can split off from the collapse  energy, different from the one-photon Rabi-Stark model in which all levels
collapse.
\end{abstract}

\maketitle

\section{ Introduction}

The most simple model in the light-matter interaction systems is the semi-classical Rabi model ~\cite{Rabi}£¬  which describes the interaction of a two-level atom and a classical light field.
If the classical light field is treated  as  a quantized mode of an optical cavity,  it was known as  quantum Rabi model (QRM). However the QRM was hardly treated  analytically  at the early stage. To solve this model readily, one employs the rotating wave approximation (RWA)~\cite{JC} so that the excitations of the atom and the photon number are conserved. Fortunately, this approximation was shown  to be very effective in the earlier cavity quantum electrodynamics (QED) system~\cite{book} at the extremely weak coupling regime for nearly resonant interaction. The basic physics realized and
observed in the earlier experiments can be described by the QRM in the RWA, such as  phenomenon of vacuum Rabi splitting,  and collapse and revivals in the population dynamics.
However,  in the past decade, with the progress of the  advanced solid devices, such as a superconducting qubit-oscillator circuit and trapped ions, the ultrastrong
coupling \cite{Niemczyk,Forn1}, even deep strong coupling \cite%
{Forn2,Yoshihara} between the artificial atom and the  quantum oscillator  have been
accessed,  the consideration of the full QRM should be highly called for. On the
other hand, the two-level system is just a qubit, which is the building block of
quantum information science and quantum computations. Just motivated by the experimental
developments and potential applications in quantum information technologies, the
QRM without the RWA has attracted extensive attentions theoretically \cite%
{Casanova,Braak, Chen2012, Chen2,Zhong,He,luo2,plenio, hgluo,Moroz}. For
more complete review, please refer to Refs. ~\cite%
{Braak2,Qiongtao,yuxi,ReviewF,Boit}.

The QRM and its variants are continuously employed to describe the physics phenomena observed in the recent experiments or quantum simulations.  A nonlinear Stark coupling between the atom and the cavity was first  generated in  an atom interacting with a high finesse optical cavity mode by applying  two laser fields
\cite{Grimsmo,Grimsmo2}. The QRM with this Stark  term was later named the
quantum Rabi-Stark model (RSM) \cite{Eckle}. This model also attracts many
attentions in recent years \cite{Eckle,Maciejewski,Xie, Xie2,Cong}. The
nonlinear Stark coupling in the RSM can induce the spectra collapse as well
as the first-order quantum phase transition (QPT).

On the other hand, the other nonlinear coupling in the form of the qubit and
two photons is also a subject of interest for a long time \cite%
{Toor,Peng,law,Emary,Dolya,Albert,Trav,Chen2012,Felicetti,duan2016,Zhangyz}.
In contrast to its one-photon counterpart, some peculiar properties can be
driven by the two-photon coupling, such as the spectra collapse. This
so-called two-photon QRM (tpQRM) becomes a hot topic recently \cite%
{Zhiguo,Maciejewski2,fanheng2,Penjie19,Felicetti1,Cong19,plenio2,Malekakhlagh,Lupo,mexi,Andrzej,Dodonov,Boas}%
. The more recent extensions include the two-photon Dicke model \cite%
{Garbe,XYchen,Alex,Jinjl,guowa} and multiphoton and nonlinear dissipative
spin-boson models \cite{Puebla2}.

Since both kinds of the nonlinear coupling can induce the spectra collapse,
what is the cooperated effect? Can the nonlinear Stark coupling induce the
first order QPT in the tpQRM? In this work, we add the Stark coupling terms
to the two-photon Rabi model, which can be named two-photon RSM (tpRSM). The
analytical solutions to tpRSM should be very intersting at the present
stage. Some crucial issue mentioned above should be addressed.

The tpRSM would be possibly realized in the solid devices such as the
trapped ions, circuit QED systems. Actually the implementation of the
one-photon RSM using a single trapped ion has been demonstrated recently
\cite{Cong}. In this proposal, external laser beams are applied to induce an
interaction between an electronic transition and the motional degree of
freedom, thus the Stark term is generated. Since the tpQRM has been realized
in the trapped ions \cite{Felicetti}, a route to simulate the tpRSM is also
experimentally feasible with a laser driving in the same device.

The paper is structured as follows: In section 2, we extend the previous
Bogoliubov operator approach to the tpRSM. A transcendent function is
derived in a \ compact way, which can give the exact regular spectra of this
model. The pole structure of the derived transcendant function is analyzed
in section 3, and thus the level crossings and the energy spectra collapse
are also discussed. The last section contains some concluding remarks.

\section{Solutions to the two-photon Rabi-Stark model}

The Hamiltonian of the tpRSM is given by
\begin{equation}
H=-\left( \frac{1}{2}\Delta +Ua^{\dagger }a\right) \sigma _{x}+\omega
a^{\dagger }a+g\left[ \left( a^{\dagger }\right) ^{2}+a^{2}\right] \sigma
_{z},  \label{Hamiltonian}
\end{equation}%
where $\Delta $ is the qubit energy difference, $a^{\dagger }$ $\left(
a\right) $ is the photonic creation (annihilation) operator of the
single-mode cavity with frequency $\omega$, $g$ is the two-photon coupling
constants, $U$ is the nonlinear Stark coupling strength, and $\sigma
_{k}(k=x,y,z)$ $\ $ are the Pauli matrices.

The Hamiltonian~(\ref{Hamiltonian}) is also invariant under a discrete
symmetry: $\hat{P}_{1}H\hat{P}_{1}^{\dagger }=H$ where $\hat{P}_{1}=\exp (-%
\frac{i\pi }{2}a^{\dagger }a)\sigma _{x}$, similar to the tpQRM. The
operator $\hat{P}_{1}$ generates the group $\mathbb{Z}_{4}$, therefore the
Hilbert space separates into four invariant subspaces. Because $\hat{P}%
_{1}^{2}=\exp (i\pi a^{\dagger }a)=:\hat{P}_{2}$, with $\hat{P}_{2}^{2}=1\!\!1$, we have also a $\mathbb{Z}_{2}$-symmetry only in the
bosonic Hilbert space. The invariance under $\hat{P}_{2}$ is reflected in
the presence of two inequivalent representations of $su(1,1)$.

On the basis of $\sigma _{z}$, the tpRSM Hamiltonian can be written in the
following matrix form in units of $\omega =1$
\begin{equation}
H=\left(
\begin{array}{ll}
a^{\dagger }a+g\left[ \left( a^{\dagger }\right) ^{2}+a^{2}\right]  &
~-\left( \frac{1}{2}\Delta +Ua^{\dagger }a\right)  \\
~-\left( \frac{1}{2}\Delta +Ua^{\dagger }a\right)  & a^{\dagger }a-g\left[
\left( a^{\dagger }\right) ^{2}+a^{2}\right]
\end{array}%
\right) .  \label{2p_RS}
\end{equation}%
We begin by performing the following Bogoliubov transformation on the
bosonic degree of freedom,
\begin{equation}
b=ua+va^{\dagger },\qquad b^{\dagger }=ua^{\dagger }+va,  \label{bogol}
\end{equation}%
where
\[
u=\sqrt{\frac{1+\beta }{2\beta }},\qquad v=\sqrt{\frac{1-\beta }{2\beta }},
\]%
with $\beta =\sqrt{1-4\gamma ^{2}}$ assumed to be nonnegative. In the tpQRM~%
\cite{duan2016}, we need set $\gamma =g$, so that one diagonal element of
the Hamiltonian matrix can be transformed into the free particle form,
facilitating the further study. In the tpRSM, due to the presence of the
Stark coupling terms, this requirement might be relaxed. We will see  that $%
\gamma $ is indeed not equal to $g$, and also depends on the nonlinear Stark
coupling strength $U$ (cf. equation (\ref{gamma}) below).

In terms of the new operator $b^{\dagger }(b)$, the upper and lower diagonal
matrix elements of the Hamiltonian can be written as
\[
H_{11}=\left( u^{2}+v^{2}-4guv\right) b^{\dagger }b-\frac{\gamma -g}{\beta }%
\left( b^{\dagger ^{2}}+b^{2}\right) -2guv+v^{2},
\]%
\[
H_{22}=\left( u^{2}+v^{2}+4guv\right) b^{\dagger }b-\frac{\gamma +g}{\beta }%
\left( b^{\dagger ^{2}}+b^{2}\right) +2guv+v^{2}.
\]%
The off-diagonal matrix elements read
\[
H_{12}=H_{21}=-\frac{1}{2}\Delta -\frac{1}{2}U\left[ \left(
u^{2}+v^{2}\right) b^{\dagger }b-uv\left( b^{\dagger 2}+b^{2}\right) +v^{2}%
\right] .
\]

The operators $b^{\dagger }b$, $(b^{\dagger })^{2}$, $b^{2}$ provide a
representation of the non-compact Lie algebra $su(1,1)$ with
\begin{equation}
K_{0}=\frac{1}{2}\left( b^{\dagger }b+\frac{1}{2}\right) ,\qquad K_{+}=\frac{%
1}{2}\left( b^{\dagger }\right) ^{2},\qquad K_{-}=\frac{1}{2}b^{2},
\label{osc-rep}
\end{equation}%
we have
\[
\left[ K_{0},K_{\pm }\right] =\pm K_{\pm },\qquad \left[ K_{+},K_{-}\right]
=-2K_{0}.
\]%
The quadratic Casimir operator $C_{2}$ of the algebra is given by
\[
C_{2}=K_{+}K_{-}-K_{0}\left( K_{0}-1\right) .
\]
In terms of the $K_{0}$, $K_{\pm }$, the diagonal elements in the
Hamiltonian read
\[
H_{11}=\frac{1-4g\gamma }{\beta }2K_{0}-\frac{\gamma -g}{\beta }%
2(K_{+}+K_{-})-\frac{1}{2},
\]%
\[
H_{22}=\frac{1+4g\gamma }{\beta }2K_{0}-\frac{\gamma +g}{\beta }%
2(K_{+}+K_{-})-\frac{1}{2}.
\]%
The off-diagonal matrix elements are
\begin{equation}
H_{12}=-\frac{\Delta -U}{2}-\frac{2U}{\beta }\left[ K_{0}-\gamma
(K_{+}+K_{-})\right] .
\end{equation}

The unitary oscillator representation \ref{osc-rep} of $su(1,1)$ in the
Hilbert space $\mathcal{H}$ generated by $b^{\dagger }$ on the state $%
|0\rangle _{\mathrm{b}}$, defined as $b|0\rangle _{\mathrm{b}}=0$, decays
into two irreducible representations, characterized by their lowest weight $%
q $: $K_{0}|q,0\rangle _{\mathrm{b}}=q|q,0\rangle _{\mathrm{b}}$. For the
even subspace $\mathcal{H}_{\frac{1}{4}}=\left\{ {b^{\dagger }}^{n}|0\rangle
_{\mathrm{b}},n=0,2,4,\ldots \right\} $, we have $q=\frac{1}{4}$ and for $%
\mathcal{H}_{\frac{3}{4}}=\left\{ {b^{\dagger }}^{n}|0\rangle _{\mathrm{b}%
},n=1,3,5,\ldots \right\} $, $q=\frac{3}{4}$. $C_{2}=\frac{3}{16}$ in both
cases. $\mathcal{H}$ separates therefore into two subspaces, $\mathcal{H}=%
\mathcal{H}_{\frac{1}{4}}\oplus \mathcal{H}_{\frac{3}{4}}$. The Bargmann
index $q$ allows us to deal with both cases independently. A basis of $%
\mathcal{H}_{q}$ is given by the normalized states
\begin{eqnarray*}
\left\vert q,n\right\rangle _{\mathrm{b}} &=&\frac{\left( b^{\dagger
}\right) ^{2\left( n+q-\frac{1}{4}\right) }}{\sqrt{\left[ 2\left( n+q-\frac{1%
}{4}\right) \right] !}}\left\vert 0\right\rangle _{\mathrm{b}}=\left\vert
2\left( n+q-\frac{1}{4}\right) \right\rangle _{\mathrm{b}}, \\
&q=&\frac{1}{4},\frac{3}{4},\qquad n=0,1,2,...\infty .
\end{eqnarray*}%
The operators satisfy
\begin{eqnarray}
K_{+}\left\vert q,n\right\rangle _{\mathrm{b}} &=&\sqrt{\left( n+q+\frac{3}{4%
}\right) \left( n+q+\frac{1}{4}\right) }\left\vert q,n+1\right\rangle _{%
\mathrm{b}},  \nonumber \\
K_{-}\left\vert q,n\right\rangle _{\mathrm{b}} &=&\sqrt{\left( n+q-\frac{1}{4%
}\right) \left( n+q-\frac{3}{4}\right) }\left\vert q,n-1\right\rangle _{%
\mathrm{b}},  \nonumber \\
K_{0}\left\vert q,n\right\rangle _{\mathrm{b}} &=&\left( n+q\right)
\left\vert q,n\right\rangle _{\mathrm{b}}.  \label{operator}
\end{eqnarray}

Note that the vacuum with respect to the original boson operators $%
a,a^{\dagger }$, $|0\rangle _{\mathrm{a}}$, with the property $a|0\rangle _{%
\mathrm{a}}=0$, may be expressed in terms of $\left\vert \frac{1}{4}%
,n\right\rangle _{\mathrm{b}}$ as
\[
\left\vert 0\right\rangle _{\mathrm{a}}=\sum_{n=0}^{\infty }z_{n}^{(\frac{1}{%
4})}\left\vert \frac{1}{4},n\right\rangle _{\mathrm{b}},
\]%
because the decomposition $\mathcal{H}=\mathcal{H}_{\frac{1}{4}}\oplus
\mathcal{H}_{\frac{3}{4}}$ is left invariant by the Bogoliubov
transformation (\ref{bogol}). We can write therefore $|0\rangle _{\mathrm{a}%
}=|\frac{1}{4},0\rangle _{\mathrm{a}}$. The condition $a\left\vert
0\right\rangle _{\mathrm{a}}=0$ gives
\begin{equation}
z_{n}^{(\frac{1}{4})}\varpropto \frac{\sqrt{\left( 2n\right) !}}{n!}\left(
\frac{v}{2u}\right) ^{n}.  \label{q1_c}
\end{equation}%
The lowest lying state (with respect to the $a$-operators) in $\mathcal{H}_{%
\frac{3}{4}}$ reads then
\[
\left\vert \frac{3}{4},0\right\rangle _{\mathrm{a}}=a^{\dagger }\left\vert
\frac{1}{4},0\right\rangle _{\mathrm{a}}=\left( ub^{\dagger }-vb\right)
\sum_{n=0}^{\infty }z_{n}^{(\frac{1}{4})}\left\vert \frac{1}{4}%
,n\right\rangle _{\mathrm{b}}=\sum_{n=0}^{\infty }z_{n}^{(\frac{3}{4}%
)}\left\vert \frac{3}{4},n\right\rangle _{\mathrm{b}},
\]%
where
\begin{equation}
z_{n}^{(\frac{3}{4})}\varpropto \frac{\sqrt{\left( 2n+1\right) !}}{n!}\left(
\frac{v}{2u}\right) ^{n}.  \label{q2_c}
\end{equation}%
In summary,
\begin{equation}
z_{n}^{(q)}\varpropto \frac{\sqrt{\left[ 2\left( n+q-\frac{1}{4}\right) %
\right] !}}{n!}\left( \frac{v}{2u}\right) ^{n}.  \label{q12_c}
\end{equation}

An eigenfunction $|\psi ,E\rangle ^{(q)}$ of $H$ with eigenvalue $E$ may be
expanded in terms of the $b$-operators as
\begin{equation}
\left\vert \psi ,E\right\rangle ^{(q)}=\left( \
\begin{array}{l}
\sum_{m=0}^{\infty }\sqrt{\left[ 2\left( m+q-\frac{1}{4}\right) \right] !}%
e_{m}^{(q)}\left\vert q,m\right\rangle _{\mathrm{b}} \\
\sum_{m=0}^{\infty }\sqrt{\left[ 2\left( m+q-\frac{1}{4}\right) \right] !}%
f_{m}^{(q)}\left\vert q,m\right\rangle _{\mathrm{b}}%
\end{array}%
\right) ,
\end{equation}%
where $e_{m}^{(q)}$ and $e_{m}^{(q)}$are coefficients to be determined
later. \ The Schr\"{o}dinger equations are given by
\begin{eqnarray*}
&&\sum_{m=0}\left[ \frac{1-4g\gamma }{\beta }2\left( n+q\right) -\frac{1}{2}%
-E\right] \sqrt{\left[ 2\left( m+q-\frac{1}{4}\right) \right] !}%
e_{m}^{(q)}\left\vert q,m\right\rangle _{\mathrm{b}} \\
&&-\frac{\gamma -g}{\beta }\sum_{m=0}\left\{ e_{m-1}^{(q)}+\left[ 2\left(
m+1+q-\frac{1}{4}\right) \right] \left[ 2\left( m+1+q-\frac{3}{4}\right) %
\right] e_{m+1}^{(q)}\right\}  \\
&&\sqrt{\left[ 2\left( m+q-\frac{1}{4}\right) \right] !}\left\vert
q,m\right\rangle _{\mathrm{b}} \\
&&-\sum_{m=0}\left[ \frac{\Delta -U}{2}+\frac{2U}{\beta }\left( n+q\right) %
\right] \sqrt{\left[ 2\left( m+q-\frac{1}{4}\right) \right] !}%
f_{m}^{(q)}\left\vert q,m\right\rangle _{\mathrm{b}} \\
&&+\frac{U\gamma }{\beta }\sum_{m=0}\left\{ f_{m-1}^{(q)}+\left[ 2\left(
m+1+q-\frac{1}{4}\right) \right] \left[ 2\left( m+1+q-\frac{3}{4}\right) %
\right] f_{m+1}^{(q)}\right\}  \\
&&\sqrt{\left[ 2\left( m+q-\frac{1}{4}\right) \right] !}f_{m}^{(q)}\left%
\vert q,m\right\rangle _{\mathrm{b}} \\
&=&0,
\end{eqnarray*}%
for the upper level, and
\begin{eqnarray*}
&&-\sum_{m=0}\left[ \frac{\Delta -U}{2}+\frac{2U}{\beta }\left( n+q\right) %
\right] \sqrt{\left[ 2\left( m+q-\frac{1}{4}\right) \right] !}%
e_{m}^{(q)}\left\vert q,m\right\rangle _{\mathrm{b}} \\
&&+\frac{U\gamma }{\beta }\sum_{m=0}\left\{ e_{m-1}^{(q)}+\left[ 2\left(
m+1+q-\frac{1}{4}\right) \right] \left[ 2\left( m+1+q-\frac{3}{4}\right) %
\right] e_{m+1}^{(q)}\right\}  \\
&&\sqrt{\left[ 2\left( m+q-\frac{1}{4}\right) \right] !}e_{m}^{(q)}\left%
\vert q,m\right\rangle _{\mathrm{b}} \\
&&+\sum_{m=0}\left[ \frac{1+4g\gamma }{\beta }(n+q)-\frac{1}{2}-E\right]
\sqrt{\left[ 2\left( m+q-\frac{1}{4}\right) \right] !}f_{m}^{(q)}\left\vert
q,m\right\rangle _{\mathrm{b}} \\
&&-\frac{\gamma +g}{\beta }\sum_{m=0}\left\{ f_{m-1}^{(q)}+\left[ 2\left(
m+1+q-\frac{1}{4}\right) \right] \left[ 2\left( m+1+q-\frac{3}{4}\right) %
\right] f_{m+1}^{(q)}\right\}  \\
&&\sqrt{\left[ 2\left( m+q-\frac{1}{4}\right) \right] !}f_{m}^{(q)}\left%
\vert q,m\right\rangle _{\mathrm{b}} \\
&=&0,
\end{eqnarray*}%
for the lower level, where the operator properties (\ref{operator}) have
been used. Projecting both sides of the Schr\"{o}dinger equation onto $_{%
\mathrm{b}}\left\langle q,n\right\vert $ yields
\begin{eqnarray}
&&\left[ \frac{1-4g\gamma }{\beta }2\left( n+q\right) -\frac{1}{2}-E\right]
e_{n}^{(q)}-\frac{\gamma -g}{\beta }\Lambda _{n}+\frac{U\gamma }{\beta }%
\digamma _{n}  \nonumber \\
&&-\left[ \frac{\Delta -U}{2}+\frac{2U}{\beta }\left( n+q\right) \right]
f_{n}^{(q)}=0,  \label{three_1}
\end{eqnarray}%
\begin{eqnarray}
&&-\left[ \frac{\Delta -U}{2}+\frac{2U}{\beta }\left( n+q\right) \right]
e_{n}^{(q)}+\frac{U\gamma }{\beta }\Lambda _{n}-\frac{\gamma +g}{\beta }%
\digamma _{n}  \nonumber \\
&&+\left[ \frac{1+4g\gamma }{\beta }2\left( n+q\right) -\frac{1}{2}-E\right]
f_{n}^{(q)}=0,  \label{three_2}
\end{eqnarray}%
where%
\begin{eqnarray*}
\Lambda _{n} &=&e_{n-1}^{(q)}+\left[ 2\left( n+1+q-\frac{1}{4}\right) \right]
\left[ 2\left( n+1+q-\frac{3}{4}\right) \right] e_{n+1}^{(q)}, \\
\quad \digamma _{n} &=&f_{n-1}^{(q)}+\left[ 2\left( n+1+q-\frac{1}{4}\right) %
\right] \left[ 2\left( n+1+q-\frac{3}{4}\right) \right] f_{n+1}^{(q)}.
\end{eqnarray*}%
If multiplying equation (\ref{three_1}) by $\frac{\gamma +g}{\beta }$ and
equation (\ref{three_2}) by $\frac{U\gamma }{\beta }$, we have%
\begin{eqnarray*}
&&\frac{\gamma +g}{\beta }\left[ \frac{1-4g\gamma }{\beta }2\left(
n+q\right) -\frac{1}{2}-E\right] e_{n}^{(q)}-\frac{\gamma +g}{\beta }\frac{%
\gamma -g}{\beta }\Lambda _{n} \\
&&+\frac{U\gamma }{\beta }\frac{\gamma +g}{\beta }\digamma _{n}-\frac{\gamma
+g}{\beta }\left[ \frac{\Delta -U}{2}+\frac{2U}{\beta }\left( n+q\right) %
\right] f_{n}^{(q)}=0,
\end{eqnarray*}%
\begin{eqnarray*}
&&-\frac{U\gamma }{\beta }\left[ \frac{\Delta -U}{2}+\frac{2U}{\beta }\left(
n+q\right) \right] e_{n}^{(q)}+\left( \frac{U\gamma }{\beta }\right)
^{2}\Lambda _{n}-\frac{U\gamma }{\beta }\frac{\gamma +g}{\beta }\digamma _{n}
\\
&&+\frac{U\gamma }{\beta }\left[ \frac{1+4g\gamma }{\beta }2\left(
n+q\right) -\frac{1}{2}-E\right] f_{n}^{(q)}=0.
\end{eqnarray*}%
Summation of the above two equation gives%
\begin{eqnarray*}
&&\frac{\gamma +g}{\beta }\left[ \frac{1-4g\gamma }{\beta }2\left(
n+q\right) -\frac{1}{2}-E\right] e_{n}^{(q)}-\frac{U\gamma }{\beta }\left[
\frac{\Delta -U}{2}+\frac{2U}{\beta }\left( n+q\right) \right] e_{n}^{(q)} \\
&&+\left[ \left( \frac{U\gamma }{\beta }\right) ^{2}-\frac{\gamma +g}{\beta }%
\frac{\gamma -g}{\beta }\right] \Lambda _{n}-\frac{\gamma +g}{\beta }\left[
\frac{\Delta -U}{2}+\frac{2U}{\beta }\left( n+q\right) \right] f_{n}^{(q)} \\
&&+\frac{U\gamma }{\beta }\left[ \frac{1+4g\gamma }{\beta }2\left(
n+q\right) -\frac{1}{2}-E\right] f_{n}^{(q)}=0.
\end{eqnarray*}%
If set%
\begin{equation}
\gamma =\sqrt{\frac{g^{2}}{1-U^{2}}},  \label{gamma}
\end{equation}%
we can remove $\Lambda _{n}$, and obtain a linear relation between the
coefficients $e_{n}^{(q)}$ and $f_{n}^{(q)}$ below
\begin{equation}
e_{n}^{(q)}=\Omega _{n}^{(q)}f_{n}^{(q)},  \label{enfn}
\end{equation}%
where
\begin{equation}
\Omega _{n}^{(q)}=\frac{(\gamma +g)\left[ \frac{\Delta -U}{2}+\frac{2U}{%
\beta }\left( n+q\right) \right] -U\gamma \left[ \frac{1+4g\gamma }{\beta }%
2\left( n+q\right) -\frac{1}{2}-E\right] }{(\gamma +g)\left[ \frac{%
1-4g\gamma }{\beta }2\left( n+q\right) -\frac{1}{2}-E\right] -U\gamma \left[
\frac{\Delta -U}{2}+\frac{2U}{\beta }\left( n+q\right) \right] }.
\label{ratio}
\end{equation}%
In either equation (\ref{three_1}) or equation (\ref{three_2}), we replace $%
e_{n}^{(q)}$ by $f_{n}^{(q)}$ through the relation (\ref{enfn}), and then
arrive at a three-term recurrence relation for $f_{n}^{(q)}$.
\begin{equation}
f_{n+1}^{(q)}=\frac{a_{n}^{(q)}f_{n-1}^{(q)}+b_{n}^{(q)}f_{n}^{(q)}}{%
c_{n}^{(q)}},  \label{rec}
\end{equation}%
where
\begin{eqnarray}
a_{n}^{(q)} &=&\frac{U\gamma }{\beta }-\frac{\gamma -g}{\beta }\Omega
_{n-1}^{(q)},  \label{an} \\
b_{n}^{(q)} &=&\left[ 2\frac{1-4g\gamma }{\beta }\left( n+q\right) -\frac{1}{%
2}-E\right] \Omega _{n}^{(q)}-\left[ \frac{\Delta -U}{2}+\frac{2U}{\beta }%
(n+q)\right] ,  \label{bn} \\
c_{n}^{(q)} &=&4\left( \frac{\gamma -g}{\beta }\Omega _{n+1}^{(q)}-\frac{%
U\gamma }{\beta }\right) \left( n+q+\frac{1}{4}\right) \left( n+q+\frac{3}{4}%
\right) .  \label{cn}
\end{eqnarray}%
All coefficients $f_{n}^{(q)}$ can be calculated with initial conditions $%
f_{0}^{(q)}=1$, $f_{-1}^{(q)}=0$.

Because $H$ is invariant with respect to the action of $\hat{P}_{1}$ in each
space $\mathcal{H}_{q}\otimes \mathbb{C}^{2}$, we can project the
wavefunction $|\psi ,E\rangle ^{(q)}$ onto $|\uparrow \rangle |q,0\rangle _{%
\mathrm{a}}$ and $|\downarrow \rangle |q,0\rangle _{\mathrm{a}}$,
respectively, to define the $G$-function as
\begin{eqnarray}
G_{\pm }^{(q)}(E) &=&\left\langle \uparrow \right\vert _{\mathrm{a}%
}\left\langle q,0\right\vert \left\vert \psi ,E\right\rangle ^{(q)}-\Pi
\left\langle \downarrow \right\vert _{\mathrm{a}}\left\langle q,0\right\vert
\left\vert \psi ,E\right\rangle ^{(q)}  \nonumber \\
&=&\sum_{n=0}^{\infty }\;\left[ f_{n}^{(q)}-\Pi e_{n}^{(q)}\;\right] \frac{%
\left[ 2\left( n+q-\frac{1}{4}\right) \right] !}{n!}\left( \frac{v}{2u}%
\right) ^{n},  \label{G-tprsm}
\end{eqnarray}%
where equation (\ref{q12_c}) has been used, $\Pi =\pm 1$, corresponding to
positive(negative) parity. The similar procedures for the derivation of the $%
G$-function within the Bogoliubov operators approach have been outlined in
the tpQRM ~\cite{duan2016,fanheng2} as well as the one-photon RSM~\cite{Xie}.

We see that each eigenstate can be labeled by three quantum numbers, the
index $n=0,1,2,\ldots $ corresponding to the continuous (bosonic) degree of
freedom, the Bargmann index $q=\frac{1}{4},\frac{3}{4}$ and the parity $\Pi
=\pm 1$, each taking two values. The numbers $q,\Pi $ label the four
(infinite-dimensional) invariant subspaces of $\mathcal{H}$, obtained from
the $\mathbb{Z}_{4}$-symmetry.

\begin{figure}[tbp]
\includegraphics[width=\linewidth]{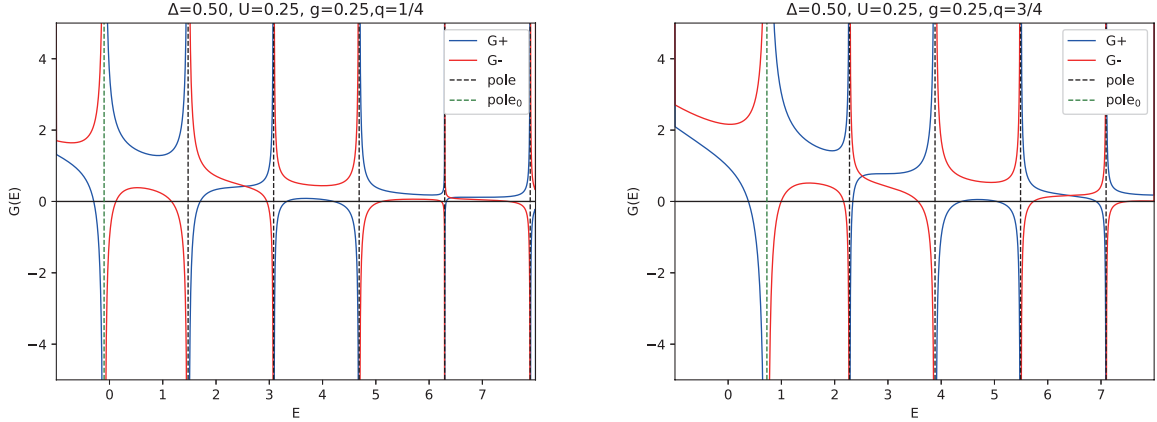}
\caption{ (Color online) $G(E)$ for tpRSM at $\Delta=0.5,g=0.25, U=0.25$, $%
q=1/4$ (left) and $q=3/4$ (right). }
\label{gfun0}
\end{figure}

The $G$-function (\ref{G-tprsm}) is a well-defined transcendental function.
The zeros of this function should give the regular spectrum. We plot the $G$%
-functions for $\Delta =0.5,g=0.25,U=0.25$ in figure \ref{gfun0}. The zeros
are just corresponding to the regular eigenenergies, which can be confirmed
by the numerical exact diagonalizations in the truncated original Fock space.

We then discuss the validity of the tpRSM G-function. Note that the maximum
value of $\gamma $ is $1/2$. According to equation (\ref{gamma}), the
maximum coupling strength is%
\begin{equation}
g_{m}=\frac{\sqrt{1-U^{2}}}{2}.  \label{limit}
\end{equation}%
Below $g_{m}$ all states are normalizable. If $U=0$, $g_{m}=1/2$, which is
consistent with that in the tpQRM~\cite{duan2016,fanheng2}. Here one can
also note that $g_{m}=0$ for $U=\pm 1$, so no any finite critical point can
be present, in sharp contrast to the one-photon RSM.

\section{Level crossings and spectra collapse}

The exact solutions in the QRM and related models can be obtained in
different ways, even superficially in the analytical way. To the best of our
knowledge, only the so-called $G$-function technique allows to analyze the
levels distribution, level crossings, and spectra collapse, because the
roots are pinched in between the poles. These interesting issues might be
hardly treated by other analytical and numerical approaches~\cite{Felicetti}%
. The $G$-function derived above indeed has the well defined pole structure.
From figure \ref{gfun0}, one can easily observe that the $G$-curves diverge
at some energy $E$, which are the typical characteristics of pole
structures. Below, we will give their positions in a closed-form.

In equation (\ref{rec}), if $c_{n}^{(q)}$ vanishes, i.e. the denominator of $%
f_{n+1}^{(q)}$ is zero, the $G$-function (\ref{G-tprsm}) diverges,
corresponding to the $n$th pole of the $G(E)$. \ So the $n>0$ pole position
is
\begin{equation}
E_{n}^{(q)pole}=2\sqrt{\left( 1-U^{2}\right) \left( 1-U^{2}-4g^{2}\right) }%
(n+q)-\frac{1+U\Delta -U^{2}}{2}.  \label{pole_n}
\end{equation}%
For the $n=0$ pole, we need inspect $e_{0}^{(q)}$ and $f_{0}^{(q)}$. In our
construction of the $G$-function, we have set $f_{0}^{(q)}=1$, so it cannot
diverge. Only $e_{0}^{(q)}$ in $G$-function (\ref{G-tprsm}) can diverge if
the denominator of $\Omega _{0}^{(q)}$ vanishes, thus we have the zeroth
pole of the $G$-function as
\begin{equation}
E_{0}^{(q)pole}=2q\sqrt{1-U^{2}-4g^{2}}-\frac{\Delta }{2U}-\frac{1-\Delta /U%
}{2}\sqrt{1-U^{2}}.  \label{pole_0}
\end{equation}%
So equations (\ref{pole_n}) and (\ref{pole_0}) compose the whole poles in
the $G$-function of the tpRSM.

\textsl{First-order quantum phase transitions:} When both the numerator and
the denominator of $\Omega _{0}^{(q)}$ vanish, we have
\begin{equation}
\beta _{c}=\frac{1-\Delta /U}{4q}.  \label{J0}
\end{equation}%
Inserting into equation (\ref{pole_0}) gives the energy without specified
parity
\begin{equation}
E_{0}^{cross}=-\frac{\Delta }{2U},
\end{equation}%
which is independent of $q$. The lowest energy levels for both parities thus
intersect at $\beta _{c}$. It is a doubly degenerate state, corresponding to
a Juddian solution \cite{Judd}. \ The level crossing of the ground-state and
the first excited state just demonstrates a first-order QPT.

By equation (\ref{J0}) we can obtain the condition for the occurrence of the
first-order QPT as
\begin{equation}
1-4q<\Delta /U<1,  \label{cgc}
\end{equation}%
because $1>\beta >0$. We write critical coupling strength at the crossing
point explicitly for later use
\begin{equation}
g_{c}=g_{m}\sqrt{1-\frac{(1-\frac{\Delta }{U})^{2}}{16q^{2}}}.  \label{gc}
\end{equation}

\begin{figure}[tbp]
\includegraphics[width=\linewidth]{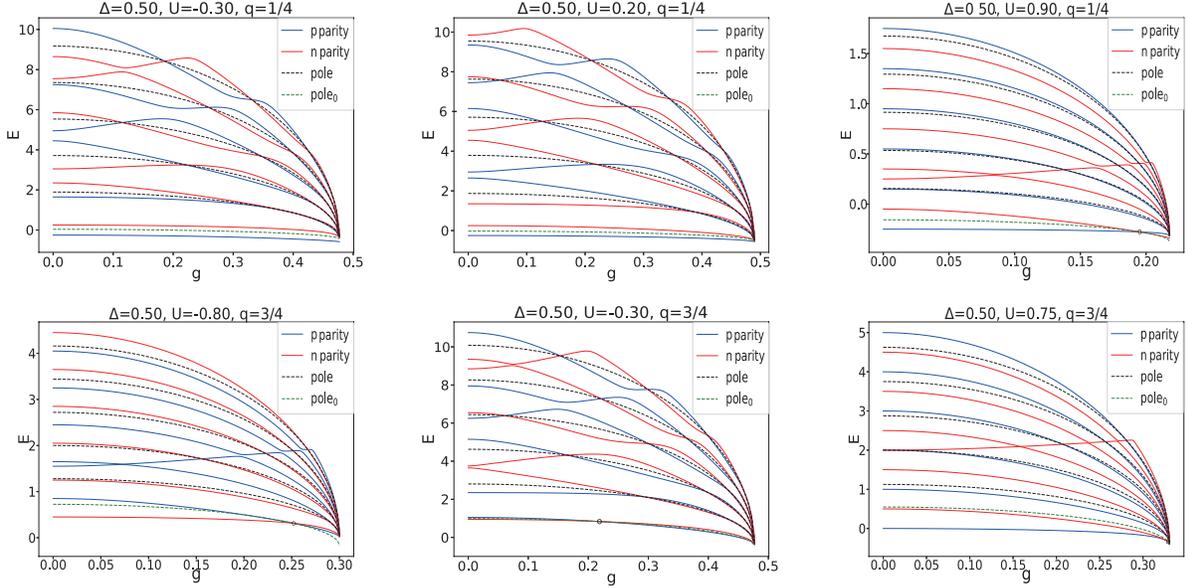}
\caption{ (Color online) Spectra for $q=\frac{1}{4}$ (upper panel) and $%
\frac{3}{4}$ (lower panel) for different values of $U$. $\Delta =0.5 $. The
first level crossing points are marked by open circles.}
\label{spectra0}
\end{figure}

To demonstrate the first-order QPT, we plot the spectra at $\Delta =0.5$ for
three typical values of $U=-0.3,0.2,0.9$ and $q=\frac{1}{4}$ in the upper
panel, for $U=-0.8,-0.3,0.75$ and $q=\frac{3}{4}$ in the lower panel of
figure \ref{spectra0}. We can see that the lowest two levels cross at $g_{c}$
exactly given by equation (\ref{gc}). For example, the predicted $%
g_{c}=\allowbreak 0.195$ for $\Delta =0.5,U=0.9,$ $q=1/4$ is exactly shown
in the upper right panel with a open circle.

\textsl{Juddian solutions for doubly degenerate states: } As usual, if both $%
c_{n}^{(q)}$ and $b_{n}^{(q)}$ in equation (\ref{rec}) vanish
simultaneously, one obtains the Juddian solutions for doubly degenerate
states. In this case, two adjacent energy levels with the positive and
negative parity can simultaneously intersect with one pole line in the
energy spectra. Below we will only discuss the special Juddian solutions
with the same crossing energy, and skip the discussions on the whole Juddian
solutions, which are in fact similar to the previous ones in both the
one-photon QRM \cite{Braak} and tpQRM \cite{Chen2012,duan2016}.

Actually, for any $n$, if both the denominator and the numerator of $\Omega
_{n}^{(q)}$ in equation (\ref{ratio}) vanish, $\Omega _{n}^{(q)}$ is
analytic, leading to analytic coefficients $e_{n}^{(q)}$ and $f_{n}^{(q)}$.
The proof is given in the following. $\Omega _{n}^{(q)}=x_{n}/y_{n}$ is
analytic for both $x_{n}=0$ and $y_{n}=0$. The denominator of $f_{n}^{(q)}$
in equation (\ref{rec}) is proportional to
\[
\left( \gamma -g\right) \Omega _{n}^{(q)}-U\gamma =\frac{\left( \gamma
-g\right) x_{n}-U\gamma y_{n}}{y_{n}}.
\]%
Thus both the denominator and the numerator of $f_{n}^{(q)}$ are also zero,
leading to analytic coefficients $f_{n}^{(q)}$, and therefore analytic
coefficients $e_{n}^{(q)}$, for the real physical systems. It can be easily
found that the limit of $\Omega _{n}^{(q)}$ in equation (\ref{ratio}) is just $-U\gamma
/\left( \gamma +g\right) $.

The condition that both the denominator and the numerator of $\Omega
_{n}^{(q)}$ in equation (\ref{ratio}) vanish simultaneously yields the
crossing coupling strength%
\begin{equation}
\beta _{c}^{n}=\frac{1-\frac{\Delta }{U}}{4(n+q)},  \label{Jn}
\end{equation}%
which  includes the first crossing point for $n=0$ discussed above.
Obviously, these Juddian solutions exist only for $\frac{\Delta }{U}<1$
because $\beta >0$ is assumed. Intriguingly, the corresponding energies at $%
\beta _{c}^{n}$ for any $n$ are the same
\begin{equation}
E_{n}^{cross}=-\frac{\Delta }{2U}>-\frac{1}{2}.  \label{cross_0}
\end{equation}%
Analogous to the one-photon RSM model \cite{Xie}, the lowest crossing energy
is also independent of the coupling constant. Here it is even independent of
the Bargmann index $q$. Surprisingly, its value is the same as that in the
one-photon RSM, indicating the independence of the detailed atom-cavity
coupling. The Juddian solutions above $E=-\frac{\Delta }{2U}$ are also
exhibited in the spectra of figure \ref{spectra0}, and are not discussed
here.

Then, we turn to the number of the crossing points situated at the same
energy $E_{n}^{cross}$(\ref{cross_0}) before a given coupling constant $g$
in the energy spectra. By equation (\ref{Jn}), we have maximum value of $n$
for the given $g$
\[
n_{\max }=\left[ \frac{1-\frac{\Delta }{U}}{4\sqrt{1-\frac{4g^{2}}{1-U^{2}}}}%
-q\right] ,
\]%
where the bracket $\left[ ...\right] $ denotes the Gaussian step function.
Including the crossing point for $n=0$, there are $n_{\max }+1$ level
crossings at the same energy $-\frac{\Delta }{2U}$ in the coupling interval $%
\left[ 0,g\right] $. The energy levels that connect these crossing points  should go  below the energy $E_{n}^{cross}=-\frac{\Delta }{2U}$ at the given $g$.
This is to say, at the coupling constant $g$ in the spectra, $n_{\max }+1$
pairs of levels are located under the energy $-\frac{\Delta }{2U}$.
Particularly, when $g\rightarrow g_{m}$, $n_{\max }\rightarrow \infty $,
there are possibly an infinite number of levels staying below the energy $-%
\frac{\Delta }{2U}$. It is very challenging to discern these crossings
without analytical reasonings outlined here, if $g$ is very close to $g_{m}$.

For $U=0.9,\Delta =0.5$, we have $g_{m}=0.21794$, $E_{n}^{cross}=-0.27778$.
If $g\leqslant 0.21$, $n_{\max }=\left[ 0.1653\right] =0$, we should have
one crossing point for $n=0$. \ This is clearly seen in the upper right
panel of figure \ref{spectra0}. If $g\leqslant 0.2178,n_{\max }=\left[
2.\allowbreak 7971\right] =2$, we should have three crossing points on a
horizontal line $\ E=-0.27778$, which is however hardly visible. The
situation becomes more serious if further approaching $g_{m}$, so the
analytical study performed above is highly called for in this case.

\begin{figure}[tbp]
\includegraphics[width=\linewidth]{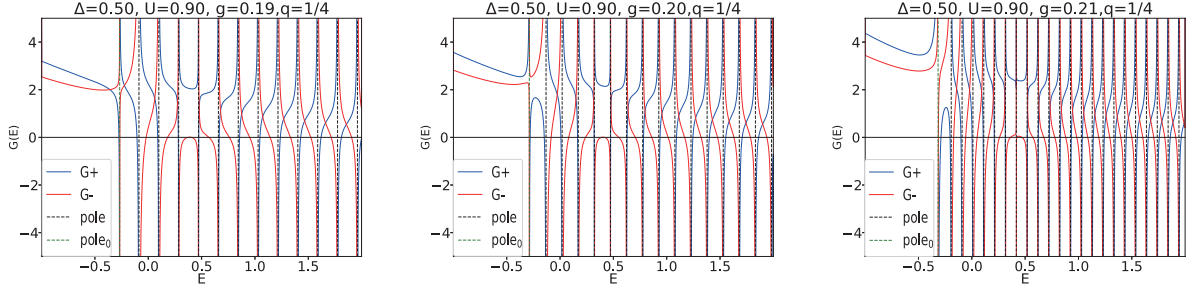}
\caption{ (Color online) $G$-curves for $g=0.19$ (left), $0.20$ (middle),
and $0.21$ (right), at $\Delta=0.5, U=0.9, q=1/4$. }
\label{gfun_U9}
\end{figure}

\textsl{Non-degenerate solutions: }  We can also discuss the other kind of the exceptional solution, so called  the non-degenerate solution~\cite%
{Maciejewski}. The non-degenerate solution is present only if one level crosses the pole line alone, which is essentially different from the  doubly degenerate solutions discussed above. One may note that the first two levels are higher than the
zeroth pole line in the upper right panel of figure \ref{spectra0}. \ For
more detail, we plot $G$-functions around $g_{c}$ in figure \ref{gfun_U9}.
For $g<g_{c}$, the first two zeros locate at the both sides of the zeroth
pole line as shown in the left $G$-curve. Just after $g_{c}$, as indicated
in the middle $G$-curve, the two zeros again go to the both sides of the
zeroth pole line in a reversal way. \ If $g$ increases further, as exhibited
in the left $G$-curve, both the first two zeros locate above the zeroth
pole. This is to say, after $g_{c}$, the level with the negative parity must
cross the zeroth pole line alone, while the level with the positive parity
does not cross at the same point. This crossing point in the energy spectra
is just corresponding to the non-degenerate solution, which can
be located in the same way as in the one-photon RSM model~\cite{Xie}.

\textsl{Spectra collapse:} For the finite Stark coupling, i.e. $U\neq 0$,
when $g\rightarrow g_{m}$, for both $q=1/4$ and $q=3/4$, all the $n>0$ poles
in equation (\ref{pole_n}) become the same%
\begin{equation}
E_{c}\rightarrow -\frac{1+U\Delta -U^{2}}{2}.  \label{energy_collapse}
\end{equation}%
Therefore all zeros of $G$-function (energy levels) are pinched in between
the poles within the vanishing intervals. It follows that they will collapse
towards the same value of $E_{c}$ when $g\rightarrow g_{m}$, which are just
demonstrated \ in figure \ref{spectra0} for all cases.

When $g\rightarrow g_{m}$, the zeroth pole described by equation (\ref%
{pole_0}) becomes
\begin{equation}
E_{0}=-\frac{\Delta }{2U}-\frac{1-\Delta /U}{2}\sqrt{1-U^{2}}.
\label{energy_co_0}
\end{equation}%
Obviously, if $U=0,$ all poles in both equation (\ref{energy_collapse}) and
equation (\ref{energy_co_0}) tend to $-1/2$, as $g\rightarrow 1/2$, which is
well known in the tpQRM. Usually the $n=0$ pole is less than $n>0$ ones. But
in the presence of the Stark coupling, one can find that $E_{0}$ is even
larger than $E_{c}$ if $0<U<\Delta $. In the limit of $g\rightarrow g_{m}$,
some energy levels could separate from the collapse energy $E_{c}$, as
exhibited in figure \ref{spectra0}. We find that the ground state energy
always separate from the collapse energy $E_{c}$ at $g_{m}$. The energy gap
does not close, indicating no continuous QPT in this model.

When $0<U<\Delta $, the collapse issue is somehow challenging. In the limit
of $g\rightarrow g_{m}$, $E_{0}$ can move above the collapse energy $E_{c}$.
\ If there are some energy levels lies in between $E_{c}$ and $E_{0}$, even
above $E_{0}$, these levels should not collapse. Since the analytical
solution at $g=g_{m}$ is still lacking at the moment, whether there are some
levels escaped from the collapse energy $\left( E_{c}\right) $ from above
remains a open question.

\section{Conclusion}

In this work, we have derived the $G$-function for the tpRSM in a compact
way by using the Bogoliubov operators approach. Zeros of the $G$-function
determine the regular spectra. The first-order QPT is detected analytically
by the pole structure of $G$-functions. The critical coupling strength of
the phase transition is obtained analytically. The occurrence of the
first-order QPT is originated from the Stark coupling, because the tpQRM
does not experience the first-order QPT. Very interestingly, the lowest
level crossing energy for $U>\Delta $ is even exactly the same as that in
the one-photon RSM, independent of the detailed dipole coupling between the
atom and cavity.

The energy spectral collapse is also found and analyzed based on the poles
of the derived $G$-function. Similar to the tpQRM, some energy levels could
escape from the collapse energy because the first pole is separated from all
other poles. The collapse characteristics is different from the one-photon
RSM where all levels collapse without exceptions, but resemble the tpQRM.
The infinite discrete upper spectra found in the one-photon RSM are absent
in the tpRSM because the terminated coupling vanishes in the limit of $%
U\rightarrow \pm 1$.

The exotic spectra observed in the tpRSM may be of fundamental importance if
the nonlinear Stark coupling is introduced to the extended two-photon
systems involving multiple levels and multiple bosonic modes, and even the
relevant open quantum systems.

\textbf{ACKNOWLEDGEMENTS} This work is supported by the National Science
Foundation of China (Nos. 11674285, 11834005).

\end{document}